# Chain Conformation and Exciton Delocalization in a Push-Pull Conjugated Polymer

Yulong Zheng,[†] Rahul Venkatesh,[‡] Connor P. Callaway,[¶] Campbell Viersen,[†] Kehinde H. Fagbohungbe,[¶] Aaron L. Liu,[‡] Chad Risko,[¶] Elsa Reichmanis,[§] and Carlos Silva-Acuña[*,†,∥,⊥]

[†]School of Chemistry and Biochemistry, Georgia Institute of Technology, 901 Atlantic Drive, Atlanta, Georgia 30332, United States

[‡]School of Chemical and Biomolecular Engineering, Georgia Institute of Technology, 311 Ferst Drive NW, Atlanta, Georgia 30332, United States

[¶]Department of Chemistry and Center for Applied Energy Research, University of Kentucky, Lexington, Kentucky 40506, United States

[§]Department of Chemical & Biomolecular Engineering, Lehigh University, 124 East Morton Street, Bethlehem, Pennsylvania 18015, United States

[∥]School of Physics, Georgia Institute of Technology, 837 State Street, Atlanta, Georgia 30332, United States

[⊥]School of Materials Science and Engineering, Georgia Institute of Technology, North Avenue, Atlanta, Georgia 30332, United States

E-mail: carlos.silva@gatech.edu

**Abstract**

Linear and nonlinear optical lineshapes reveal details of excitonic structure in semiconductor polymers. We implement absorption, photoluminescence, and transient absorption spectroscopies in DPP-DTT, an electron push-pull copolymer, to explore the relationship between their spectral lineshapes and chain conformation, deduced from resonance Raman spectroscopy and from *ab initio* calculations. The viscosity of precursor polymer solutions before film casting displays a transition that suggests gel formation above a critical concentration. Upon crossing this viscosity deflection concentration, the lineshape analysis of the absorption spectra within a photophysical aggregate model reveals a gradual increase in interchain excitonic coupling. We also observe a red-shifted and line-narrowed steady-state photoluminescence spectrum, along with increasing resonance Raman intensity in the stretching and torsional modes of the dithienothiphene unit, which suggests a longer exciton coherence length along the polymer-chain backbone. Furthermore, we observe a change of lineshape in the photoinduced absorption component of the transient absorption spectrum. The derivative-like lineshape may originate from two possibilities: a new excited-state absorption, or from Stark effect, both of which are consistent with the emergence of high-energy shoulder as seen in both photoluminescence and absorption spectra. Therefore, we conclude that the exciton is more dispersed along the polymer chain backbone with increasing concentrations, leading to the hypothesis that the polymer chain order is enhanced when the push-pull polymers are processed at higher concentrations. Thus, tuning the microscopic chain conformation by concentration would be another factor of interest when considering the polymer assembly pathways for pursuing large-area and high-performance organic optoelectronic devices.

# INTRODUCTION

The most fundamental aspect of the materials science of semiconductor polymers concerns the relationship between solid-state microstructure, macromolecular conformation, and the electronic and optical properties of these materials. From a molecular perspective, the optical properties of organic semiconductors are governed by Frenkel excitons,[1] and this relationship entails the dependence of the exciton optical transition density on the conformation of the polymer backbone, and on the nature of Coulomb coupling between chain segments, both intra- and interchain.[2] The HJ or photophysical aggregate models, which account for a collective ensemble of interacting transition dipole moments developed by Spano have been spectacularly successful in quantifying excitonic coupling and the nature of the disordered energy landscape from linear spectral lineshapes.[3–10] These models have been extended to push-pull polymers,[11–13] in which the role of charge-transfer interactions are evident. In this article, we examine absorption, photoluminescence (PL), and transient absorption optical lineshapes in films of poly[2,5-(2-octyldodecyl)-3,6-diketopyrrolopyrrole-*alt*-5,5-(2,5-di(thien-2-yl)thieno-[3,2-b]-thiophene)] (DPP-DTT), a push-pull polymer designed for applications in thin-film transistors.[14,15] The films are cast from solutions of various concentrations; we find that the optical lineshapes of the films show a strong dependence on the precursor concentration, and correlate the spectral shapes with chain conformation derived from resonance Raman measurements and the corresponding *ab initio* calcuations. We conclude that when solutions are cast from gel-like solutions, excitons are more highly delocalized along the polymer backbone, driven by more planar, less torsionally disordered backbones.

A subtle control of the polymer aggregation could tune the polymer chain conformations, which determines the arrangement of the chromophores. It is crucial to build the correlations between the microscopic conformations and photophysical properties. Although previous studies have shown that photophysical and electronic properties of conjugated homopolymers, like poly-(3-hexyl)thiophene (P3HT) or polyphenylenevinylenes, are greatly influenced by the long-range order,[7,16–18] where the exciton dissociation and charge migrations are impacted by the polymeric aggregate fractions, microstructural conformations, defects and etc.,[19–21] the photophysical aggregates and nonaggregates of short-range order could also play a role in the photogenerated charges.[22,23] Furthermore, in P3HT-like derivatives, the exciton coherence lengths, which indicates the spatial span of the exciton wave, are varied with different molecular weights.[16] Specifically, the intrachain exciton extending more units along the chain backbone in the high molecular-weight P3HTs give rise to longer PL lifetime, compared to their low molecular-weight counterparts.[17]

In comparison to conjugated homopolymers, the photophysical properties and charge transport behavior might depend on the short-range order more subtly in electron push-pull or donor-acceptor (DA) copolymers.[24–27] Previous work showed that the electron push-pull nature enhances exciton-exciton annihilation, which gives rise to long-lived bound charge pairs.[28] Besides, intra- and interchain charge-transfer interactions, representing the wavefunction overlaps between the electron-sufficient and -deficient moiety along and across the polymer chains, could be subject to the polymer chain backbone orders and $\pi$-$\pi$ stacking, respectively.[12,24,29,30] Furthermore, the charge-transfer character in the electron push-pull polymers renders a permanent dipole and induces a strong overlap of the electron cloud between push and pull chromophores, which breaks the Kasha approximation of only transition dipole moments interacting.[31] Therefore, a different energetic landscape from that of homopolymers polymers might be expected in electron push-pull materials.

The impact of inter- and intrachain interactions on the photophysical responses of conjugated polymers has been studied through many ways, such as embedding the diluted semiconductor polymers in inert matrices, or surrounding polymer chains with molecular crowns.[32–37] In either way, the interchain excitonic interactions are considered to be reduced. However, most studies introduced new components which might lead to ambiguity regarding polymer chain order, microstructures, dielectric environment and unwanted bath-system interactions. Induced gel formation, on the other hand, only exploited either molecular weight- or concentration-dependence, which are intrinsic properties of polymers in their solution and solid state.[38,39] Here, we contribute a detailed understanding of the impact of photophysical aggregation on exciton delocalization as well as chain planarization through gel formation process in push-pull polymers.

## RESULTS

Using the electron push-pull polymer, as the material of interest displayed in Fig. 1a, DPP-DTT, thin films were prepared from solutions below and above the viscosity deflecting concentration, $c^*$ as shown in Fig. 2a. Below $c^*$, the polymer chains or preassembled small aggregates are more isolated, while they start to pack together with increasing polymer concentrations. By leaving the solutions standing still for a few days, the solutions below $c^*$ are still fluidic while the ones above $c^*$ become immobile as shown in Ref. 39 Fig. S2 The film preparation method was described previously.[40] To account for the perturbation of the Coulombic interactions within the excitation band, the absorption spectra were fit to a Franck-Condon (FC) progression modified by the contributions of exciton bandwidth, where the effective Huang-Rhys parameter was set to be 0.73 (Fig. 1b).[13] With an increase in concentration, the ratio of $A_{0-0}$ and $A_{0-1}$ absorption peaks decreases, which corresponds to an increase in interchain exciton bandwidth from 16 to 54 meV as shown in Fig. 2b, accompanied with a small blue shift around 15 meV of the $A_{0-0}$ peak (see Table S1 and Fig. 1b). The magnitude of the exciton bandwidth falls well under the weakly-coupled HJ-aggregate limit, i.e. the estimated excitonic interactions are much smaller than the reorganization energy of the ring stretching mode.[7] A direct consequence of larger excitonic interaction in the H-aggregate is an increasing Stokes shift when examining the steady-state photoluminescence (PL) and absorption measurements simultaneously (Fig. 1b). Interestingly, in addition to the redshift of the PL peaks, a trend of linewidth narrowing is also observed as shown in Fig. 2d, where the major peak can be simply fitted with a standard Gaussian distribution. Although enhanced interchain excitonic interaction could lead to the redshifting behavior in the emission of H-aggregate, it will not result in drastic line narrowing in the PL line shapes. Based on the model developed by Knapp,[41] Knoester[42] and Spano,[6,10] the line narrowing effects seen in aggregate PL spectra can be explained by the motional narrowing effect, where the distribution of static disorder is averaged out due to the fast-moving excitons. Under Kasha's approximation, the excited excitons will migrate and are most probable to be localized at the deepest traps. Therefore, the steady-state PL lineshapes would follow the distribution of the deepest traps. Specifically, the distribution of the deepest traps could be impacted by the few following physical factors; the magnitude of static disorder, (inhomogeneous broadening linewidth, $\sigma_d$), aggregate size, (number of chromophores in each aggregate, $N$), and/or exciton coherence length across chains, in the weakly-coupled H aggregate.[6,10,16] We interpret these observations as a weakly varying width of total disorder with processing concentrations. The relevant length scales for the photophysical aggregate are highly microscopic (of order nearest molecular neighbor), while X-ray crystallography sample length scales are relevant to crystalline order. Nevertheless, we can compare this information with that derived from grazing-incidence wide-angle X-ray scattering (GIWAXS) measurements. Recent measurements on this material[39] demonstrated invariant d-spacing values for $\pi - \pi$ stacking of 3.7 Å and lamellar spacing of $19.6 \pm 0.2$ Å, consistent with other DPP-based copolymers as shown in previous literature.[30,38,43] Furthermore, consistent full width at half maximum (FWHM) linewidths for each (010) scattering peak when comparing all samples, indicating similar paracrystalline static disorder among all samples.[44] To compare the aggregate sizes of the thin films for different concentrations, differential scanning calorimetry (DSC) was performed as shown in Fig. S3. A melting temperature of $376 \pm 1$ °C is found among all samples, and a consistent curve shape for the melting peak is observed. Combining both GIWAXS and DSC measurements, we deduce that the aggregate size and the lattice disorder are probably not the dominant factors for the drastic red shift as observed in the steady-state PL spectra. Since the intra- and interchain exciton delocalization are countering each other,[16,45] the motional narrowing effect could be ascribed to the variation of the exciton dispersion due to the change of polymer chain order in the HJ-type photophysical aggregates, as will be demonstrated later on.

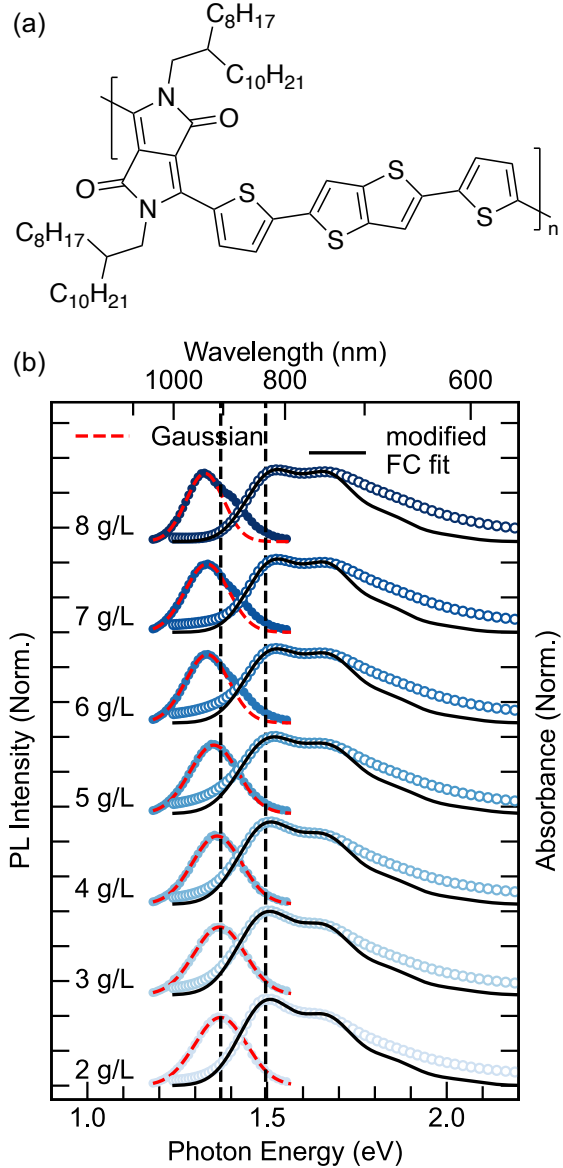

**Figure 1:** (a) Molecular structure of DPP-DTT. (b) Normalized absorption spectra (open circle), and steady-state PL (solid circle) of thin films deposited from different solution concentrations. We note that the abosption spectra were previously published in Ref. 40. The PL peaks are fitted with a single Gaussian distribution (red dash line), while the absorption vibronic replica are simulated with modified Franck-Condon progression (solid black line). The spectra shift for PL and absorption are denoted with the dashed black lines with increasing concentrations. A pronounced red shift can be readily observed in the PL spectra, meanwhile the absorption spectra also display a small blue shift of around 15 meV (see Table S1 in SI for quantitative results.)

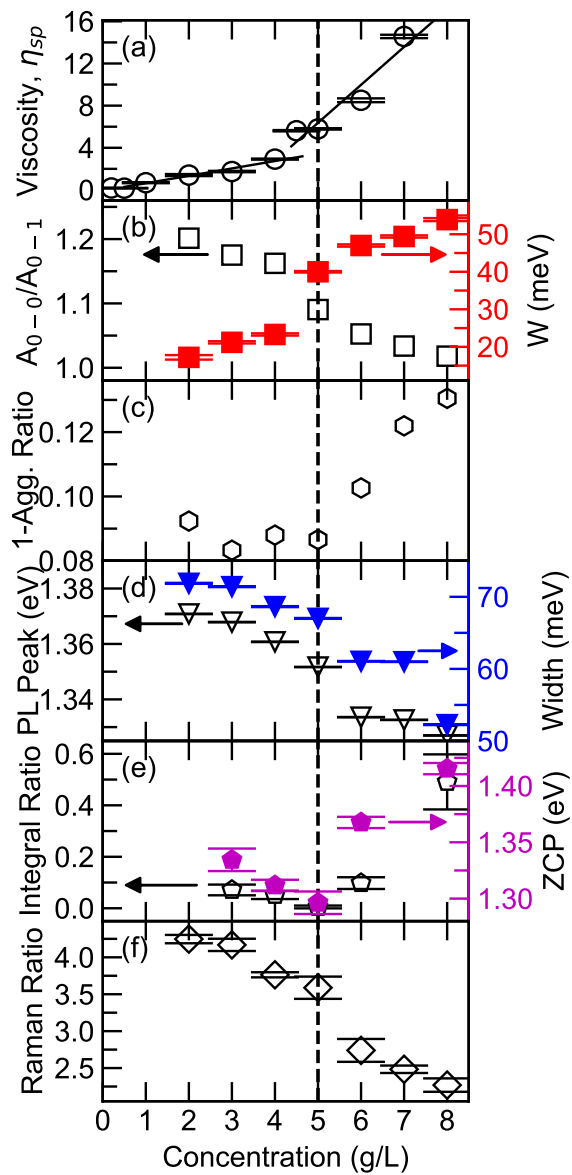

**Figure 2:** (a) Viscosity measurement on DPP-DTT-chlorobenzene solutions performed at 56°C, reproduced from Ref 40. The two-regime behavior is visualized by the two linear lines with different slopes (b) The $A_{0-0}/A_{0-1}$ ratio (black open squares) and the effective interchain exciton bandwidth (red solid squares) acquired from FC simulations as a function of concentrations. The error bars associated with exciton bandwidths are also indicated. (c) The ratio of the optical absorption area of the aggregate and nonaggregate spectra as shown in Fig. S1(open hexagon). (d) The peak positions (black open triangles) and widths (blue solid triangles) acquired from Gaussian distributions. (e) The integral ratio (open pentagons) of the shaded area in PIA and GSB, as shown in Fig. 4. The zero cross points (ZCP) are denoted in purple solid pentagons. (f) The ratio of resonance Raman peak at 1411 and 1366 cm$^{-1}$, with all spectra simultaneously normalized at peak 1525 cm$^{-1}$. The dashed line is the guide for eye of the critical concentration point.

Another distinctive feature from the PL spectra is the emerging side peak on the high energy shoulder when the concentration crosses $c^*$. As it only separated from the dominant Gaussian peak by 100 meV, we exclude the possibility of it being a vibronic satellite of the typically dominant 160-180 meV modes. A similar trend is also observed within the absorption spectra (Supplemental Fig. S1); the deviation from the aggregate spectra on the high-energy side (from 1.43 to 1.68 eV) is partially ascribed to the absorption of the non-aggregates.[7,43] Although the ratio of the extinction coefficients between the aggregates and non-aggregates are unknown, we can still deduce the change of the component ratio by assuming the molar extinction coefficients ratio being constant. As there is an increase in the optical absorption from the non-aggregates to the aggregates spectra, the relative amount of non-aggregates are supposed to increase. A constant ratio is observed before c* while the relative contribution from the non-aggregates starts to increase drastically after the critical concentration. Therefore, the new PL side peaks can be correspondingly assigned to the emission of such photophysical non-aggregates.

The contributions of the non-aggregate were probed via transient absorption (TA) spectroscopy across the viscosity deflecting concentration. In Fig. 3 we display TA spectra for 3, 4, 6 and 8 g/L films, with the pump pulse centered at 760 nm, and with fluence under 5 $\mu$J/cm$^2$. (Measurement for films prepared from 5 g/L solutions are presented in Fig. S4 and 6 in the Supplementary Information.) We note that local heating of the laser beam could lead to a change in the microstructure of the aggregates. Within the fluence range, we observe consistent lineshapes as shown Fig. S6. However, the exact effects of the local heating on the microstructural changes are worth further investigation in this material system. The four TA plots show two domains where the red and blue represent the ground state bleaching (GSB) and photoinduced absorption (PIA), respectively (Fig. 4). All subplots show a dominant GSB signal and the high absorption coefficient is commonly observed due to the long persistence length of the conjugated copolymers.[46] The lifetimes of GSB are around 18 ps (see SI Fig. S7). The PIA signals, on the other hand, show a comparable lifetime to that of the GSB, which suggests that the high-lying excited states relax to the lowest excited vibronic state on a comparable timescale. It is worth mentioning that since we adopted relatively low pump fluences here, the signals become less clean when the time delay is beyond 100 ps, which is specifically emphasized in Fig.3(d). However, as shown in Figure S5, the spectral cuts at longer time delays (beyond 100 ps) show noisy features most likely due to the low fluences we applied. It is unknown at this point if some of the features are real but further investigations on possible thermal photo products will be interesting.

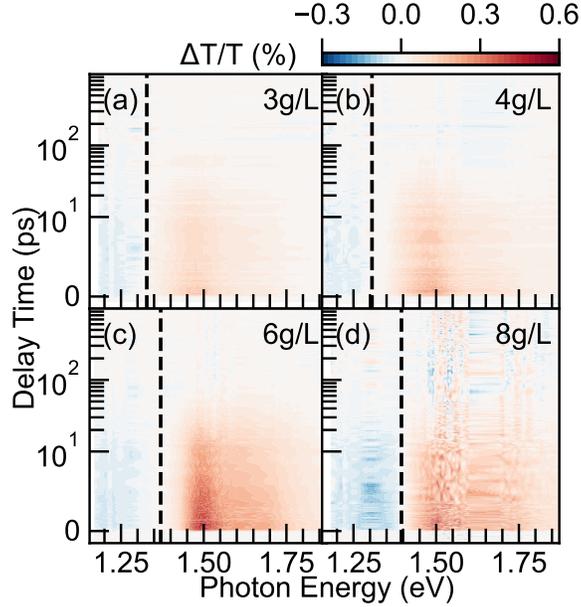

**Figure 3:** Transient absorption spectral maps showing the differential transmission signals of (a) 3, (b) 4, (c) 6, and (d) 8 g/L samples pumped with 760-nm pulsed beam under 5-$\mu$J/cm$^2$ fluence. The transmission PIA and GSB signals are colored in blue and red, respectively. The dash line lies in the zero-amplitude cross points for each sample, where a clear shift is observed.

To avoid arbitrary spectral drifting at early times, the spectra were averaged by taking the temporal cuts from 0.15 to 1 ps as displayed in Fig. 5. Within the polymer films deposited from higher concentrations, a new excited-state feature emerges in the PIA region. The zero-cross points, defined as photon energies where the differential transmission signal is zero, indicate counteracting contributions between positive signal (GSB) and negative signal (PIA). The zero cross points (black circles) with respect to the concentration are plotted in Fig. 2e. The zero-cross points are observed to have a slight decrease up to $c^*$ followed by a drastic increase. The measurement of 5-g/L film shows a clear small peak around 1.33 eV (932 nm) due to stimulated emission (SE), which also gives rise to the red shift of the zero-cross points in addition to GSB. As noted, the zero-cross points for samples of 4 and 6 g/L might also have contributions from SE, even though they are much weaker than 5-g/L sample. As the precursor's concentration increase, the new growing PIA feature starts to mask the SE signal as they are located at the same wavelength. The overlaps between SE and PIA result in such unusual behavior of the shift of zero-cross points. The relative integral ratio of the PIA feature and the GSB of the excitons shows a similar trend by comparing the integral of the absorption in these two regions (shown as the shaded area in Fig. 5a). A direct absorption from the high-energy excited states in the non-aggregates could contribute to this enhanced PIA signature, which is consistent with a growing content of non-aggregates in samples prepared from high concentrations as mentioned earlier. As mentioned above, the turning point is also displayed around the critical concentration, $c^*$, from where there is an increase in the optical absorption of the non-aggregates content. Therefore, a direct absorption from the excited states in the non-aggregates could contribute to this enhanced PIA signature, which is consistent with a growing content of non-aggregates in samples prepared from high concentrations. On the other hand, it is also worth pointing out that the new PIA features resemble the derivative-like Stark effect features,[47,48] as the differential spectra are perturbed by the induced electric field from the accumulating photogenerated charges at the interfaces of aggregates and non-aggregates.[22] The degenerate states are lifted by the induced electric field to give such new feature, especially when comparing the differential line shapes with the linear transmission spectra. Below $c^*$, the GSB transition very much follows the linear transmission, while above $c^*$, the differential spectra have a sharper transition. The two above mentioned factors could both contribute to the new PIA features.

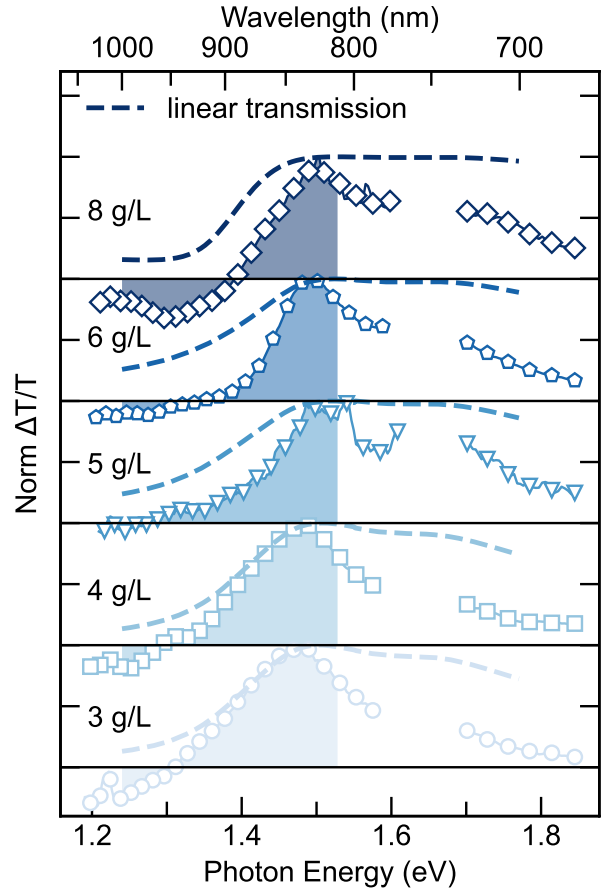

**Figure 4:** Normalized transient absorption spectra integrated from 0.15 to 1.0 ps measured for samples of 3, 4, 5, 6, and 8 g/L. The cutoff from 1.57 eV to 1.71 eV is due to the leak of pump beams. The shaded area is integrated to estimate the ratio between GSB and PIA. The differential transmission spectra are overlapped with the linear transmission spectra (dashed line) converted from Figure 1.

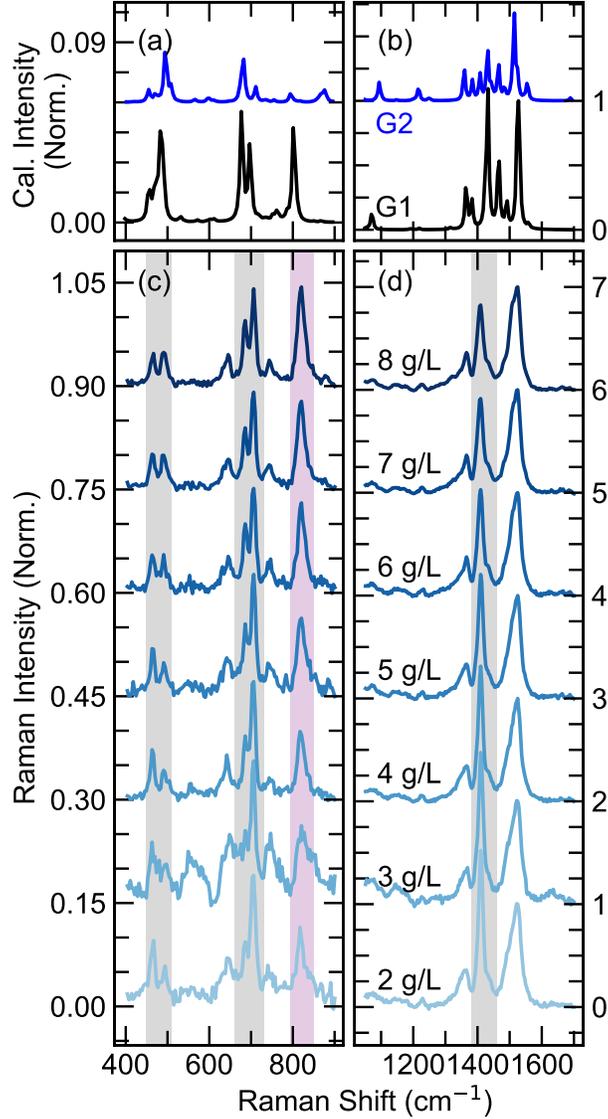

**Figure 5:** (a) and (b) show the calculated resonance Raman spectra with geometry 1 (black) and geometry 2 (blue), respectively; both are normalized against the geometry 1 peak intensity at 1525 cm$^{-1}$. The spectra are normalized by their relative calculated maximum intensity. (c) and (d) demonstrated the experimental Resonance Raman spectra displayed within both low and high Raman shift ranges. All experimental spectra are simultaneously normalized at 1525 cm$^{-1}$, which is a C=C stretching mode localized in the DPP unit. The peaks shaded in grey indicate a decrease in Raman intensity, while the peaks shaded in purple indicate increasing Raman intensities with concentrations.

To further understand the polymer chain order and local conformations, we performed resonance Raman spectroscopy to study the on-chain vibrational modes, which are coupled to the electronic transitions. As a more torsionally ordered polymer chain backbone could sustain a longer exciton coherence length,[16] a more delocalized electron density would weaken the resonance Raman intensity.[17,49] Here, the high-energy band in DPP-DTT at 488 nm was excited, which contributes to a delocalized $\pi - \pi^*$ transition as demonstrated by Wood *et al*.[49] The Raman spectra at high and low frequency range are shown in Fig. 5c and d, respectively. Interestingly, we observed no shifts of the Raman modes, while the intensities of certain modes vary substantially. The most significant change is observed at $1410\,\mathrm{cm}^{-1}$ within the high-frequency range from 1200 to $1600\,\mathrm{cm}^{-1}$ displayed in Fig. 5c, whereas the more substantial variations occur in the low-frequency range as shown in Fig. 5d.

The Raman-active modes are assigned based on density functional theory (DFT) calculations. These calculations were performed on a DPP-DTT trimer, truncated with a fourth DPP-thiophene (T) unit. The additional DPP-T unit is utilized to allow for a symmetric charge distribution. The calculated Raman spectra are displayed in Fig. 5a and b. As demonstrated by Chaudhari *et al.*, two different geometries might coexist in the ensemble, both contributing to the Raman cross section. In geometry 1 (G1), the oxygen atoms of the DPP unit are oriented close to the sulfur atoms of the neighboring T units, while in geometry 2 (G2), they are instead oriented near the hydrogen atoms of the neighboring T. The optimized structures of G1 and G2 are shown in Fig. S8. Based on DFT calculations, the ground-state energy of G1 is lower than that of G2 by approximately $41.2\,\mathrm{kJ\,mol}^{-1}$. The calculated Raman spectra of both geometries are shown in Fig. 5; both spectra are normalized to the peak at $1525\,\mathrm{cm}^{-1}$ in G1. In the following discussion, all Raman shifts refer to experimental results unless specified otherwise. A complete assignment of the Raman modes is presented in Table 1.

To allow us to compare the change of Raman intensities quantitatively among different samples, a vibrational mode that is not significantly influenced by the torsional order of the polymer backbone (i.e., a local or intraunit vibrational mode) is used as a benchmark. Herein, such mode is chosen to be the local C=C stretching in the highly rigid DPP unit at $1525\,\mathrm{cm}^{-1}$, coupled with a local, asymmetric DTT ring deformation, as shown in the vector diagram in Fig. 6a. Such localized C=C stretching mode on the DPP unit is also observed in a related

DPP-based copolymer.[49] Another localized stretching mode at $1366\,\text{cm}^{-1}$ associated with the two bridgehead carbon atoms on the DPP units, which shows constant intensity among all samples, supporting our rationalization. In contrast, the greatest changes in intensity are observed at $1416\,\text{cm}^{-1}$, where the DTT ring demonstrates strong deformation as shown in Fig. 6b. Interestingly, the corresponding simulated Raman peaks are calculated to be doubly degenerate, with one mode being the ring deformation of the second and third DTT unit and the other mode being that of the first and fourth DTT unit. Such 'globally' delocalized DTT ring deformation directly contributes to a more delocalized exciton wave along the polymer chain backbone, leading to a weaker vibronic coupling strength and decrease of the Raman intensity. A more quantitative demonstration is displayed in Fig. 2f, where the intensity ratio of Raman peaks at 1410 and $1366\,\text{cm}^{-1}$ is shown to decrease, with clear two-regime behavior.

The other two globally delocalized vibrational modes, at 706 and $467\,\text{cm}^{-1}$, show a similar decreasing trend in intensity, where the former mode is a gentle ring breathing motion among all DPP and DTT units and the latter is a degenerate ring deformation mode in all DTT units (Fig. 6f). Aside from the stretching modes, the dynamically disordered thienothiophene unit also experiences a strong torsional motion as observed at $494\,\text{cm}^{-1}$ (Fig. 6e). Besides the two different trends and their associated vibrational modes, an intriguing increasing trend is observed at $810\,\text{cm}^{-1}$, colored in purple in Fig. 5c. This mode corresponds to the global symmetric C-N stretch in the DPP units (Fig. 6d), which localizes the exciton wave. Therefore, when the polymer chain backbone becomes more planarized, which sustains a more delocalized exciton, the vibrational mode at $810\,\text{cm}^{-1}$ is not only able to collect the electron density, but also redistribute that density in an almost perpendicular direction relative to the polymer chain. The final result is an increase in intensity for higher concentration samples.

Such observation of the Raman intensity variations aligns well with the observation of the PL motional narrowing effect, as mentioned above, indicating improved polymer chain planarization in samples prepared higher than the gel formation concentration. Besides, the detailed analysis of the absorption and PL spectra show the increased exciton bandwidth, indicating stronger interchain interaction. Another surprising finding of higher optical contributions from the non-aggregates with increasing concentrations surfaces when comparing the lineshape and oscillator strength from the linear absorption and nonlinear TA measurements. The strong correlations between the viscosity and spectroscopic results can now be well established (Fig. 2).

**Table 1:** Comparison between the experimental and simulated resonance Raman modes. Simulation results correspond to DPP-DTT geometry 1.

| Exp. Raman shift (cm$^{-1}$) | Simulated Raman shift (cm$^{-1}$) | Norm. intensity 2 (g/L) | Norm. intensity 8 (g/L) | Qualitative assignment |
|---|---|---|---|---|
| 1525 | 1531 | 1$^a$ | 1$^a$ | L: 1,4-DPP $\nu_{b-C_t=C_s}$, L: asym. $\nu_{C_s=C_s}$ on 1,3-DTT |
|  | 1525 |  |  | L: 2,3-DPP $\nu_{b-C_t=C_s}$, L: sym. ring deformation on 2-DTT |
| 1491 | 1493 | 0.507±0.014 | 0.363±0.009 | D: gentle DTT and DPP rings deformation, asym. m-Th ring deformation |
| 1410 | 1431.9, 1432.0$^b$ | 1.507±0.007 | 0.841±0.023 | D: 1,2,3-DTT rings deformation due to $\nu_{b-C_t=b-C_t}$ |
| 1366 | 1383 | 0.355±0.004 | 0.371±0.009 | L: 2,3-DPP $\nu_{b-C_t=b-C_t}$ |
|  | 1381 |  |  | L: 4-DPP $\nu_{b-C_t=b-C_t}$ |
| 818 | 802 | 0.103±0.011 | 0.135±0.001 | D: sym. $\nu_{C_p-N}$ on 1,2,3,4-DPP units |
| 706 | 698 | 0.190±0.010 | 0.141±0.004 | D: gentle ring breathing of 1,2,3-DTT unit and 2,3-DPP unit. |
| 686 | 678 | 0.095±0.008 | 0.096±0.001 |  |
| 643 | — | 0.072±0.003 | 0.038±0.001 | — |
| 494 | 482 | 0.045±0.010 | 0.048±0.004 | D: torsion of 1,2,3-DTT units and ending m-Th units |
|  | 471 |  |  |  |
| 467 | 455, 456$^b$ | 0.085±0.010 | 0.044±0.004 | D: ring deformation of 1,2,3-DTT units due to the stretching of S atoms |

(i)$^a$The whole spectrum is normalized at 1525 cm$^{-1}$; $^b$Essentially degenerate modes;
(ii) L: local, D: delocalized; 1, 2, 3 and 4 label the order for DPP or DTT unit displayed in Figure 6;
(iii) $C_t$, $C_s$, $C_p$ are ternary, secondary and primary carbon atom, respectively; b-C: bridgehead carbon of polycyclic rings; m-Th: monomeric thiophene ring.

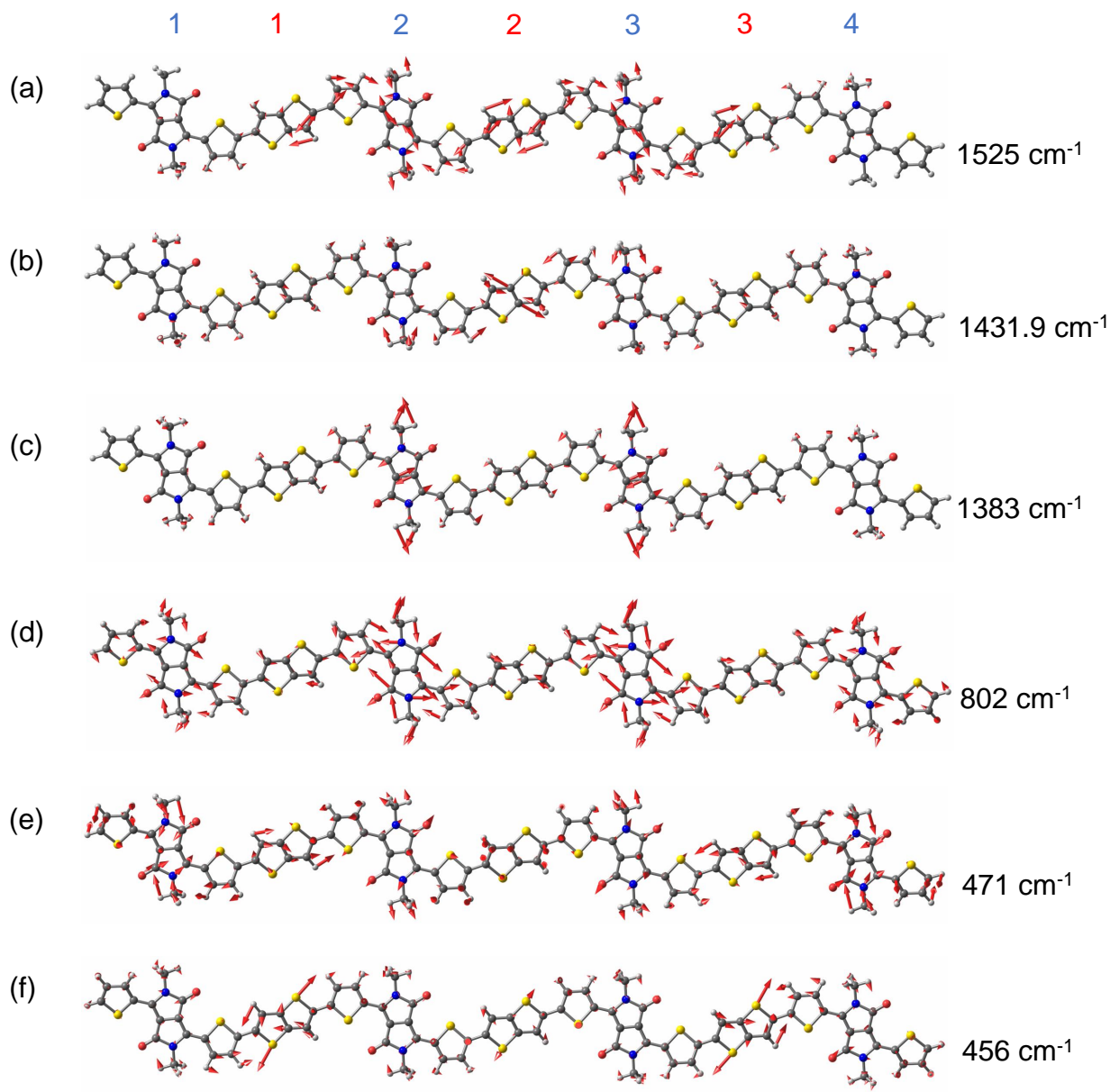

**Figure 6:** Vector diagrams for the Raman modes of a symmetrically truncated DPP-DTT trimer in geometry 1. (Raman modes for geometry 2 are shown in Figure S9.) Alkyl side chains are substituted with methyl groups for computational efficiency. Atomic color scheme: gray = carbon, white = hydrogen, yellow = sulfur, blue = nitrogen, red = oxygen. The blue and red numbers at the top indicate the indices of the DPP and DTT units, respectively. Atomic displacements are indicated by the arrows.

# DISCUSSION

Electron push-pull polymers, unlike conjugated homopolymers, exhibit strong charge-transfer character between the electron sufficient and deficient unit within chains. This intra-chain interaction plays a dominant role in the optical spectral structure of the material, particularly in enhancing the intrachain (J-like) coupling of chromophores versus inter-chain (H-like) coupling. The spectral features that arise from J-like coupling, namely the enhancement of the 0–0 spectral peak intensity in the absorption and PL vibronic progression, are attributed to the in-phase electron and hole integrals, and are prevalently shown in other electron push-pull polymers.[46,50–53] As further demonstrated by Chang *et al.*, the dominant interchain charge transfer interaction in DPP-based systems could lead to abnormal red-shifted H-type aggregate behavior, when the electron and hole transfer integral is out-of-phase, which has been shown experimentally.[43] While it can be established that the sign of charge-transfer integrals is sensitive to the donor/acceptor stacking arrangement, i.e., $\pi$–$\pi$ stacking geometry and distance,[54] the investigation of the stacking arrangement on a molecular/atomic level requires rigorous two-dimensional solid-state nuclear magnetic resonance spectroscopic studies, which to date have been limited for push-pull polymers especially in thin films.[30,55–57] Of particular relevance, Chaudhari *et al.* showed that DPP-DTT polymer aggregates adopt a geometry where the donor and acceptor units between chains are alternating stacked when the films are prepared by spin coating. However, they also point out that segregated donor-on-donor or acceptor-on-acceptor stacking arrangement might have slightly lower energy compared to the slip-on geometry. Such stacking order could be achieved when certain processing conditions are met. We posit that the fraction of chromophores that adopt such arrangements varies strongly processing conditions and with a resulting microstructure that is responsible for the trends with concentration reported in Fig. 2.

Under the influence of solution concentration, the resonance Raman spectra displayed the most changes in the low-frequency range, where most of the contributions originate from the stretching and torsional modes of the DTT unit. Such conformational disorder stems from the rotational invariance of the mesogenic groups (i.e. thienothiophene component), and the favorable lowest-energy backbone conformation could be adopted easily.[58] Indeed, previous DFT calculations performed on DPP-DTT have shown that the different conformations of the DTT unit (e.g. different twisting angles between the thienothiophene and monomeric thiophene rings) have more minor energy difference, compared to that of the DPP unit, implying that the disorder source possibly originates from the former.[30] In contrast, the DPP units are highly planarized and rigid, verified by the localized vibrational modes seen in Fig. 5d in the high-frequency region. Note that whether the Raman intensities increase, decrease, or remain constant depends on the nature of each vibrational mode; some of the thienothiophene ring deformations along the chain backbone will aid in dispersing the exciton wave, while specific DPP ring deformation modes will lead to exciton wave localization, perpendicular to the polymer chain backbone. This direction dependency reflects the tensor aspects of the Raman polarizability. The large repeat unit will have not only diagonal elements but also off-diagonal elements in the polarizability expression.

Current studies do not allow us to quantitatively determine the spatial correlation of site energies, which can be quantified by the $I_{0-0}/I_{0-1}$ PL ratio,[16] specifically, due to limitations in the PL detection range. However, as mentioned earlier, the 0-0 peaks become more red-shifted and narrowed with increasing concentration. Furthermore, from the resonance Raman spectra, we are able to conclude that a more dispersed exciton wave is achieved along the polymer chain backbone with increasing concentration. Therefore, it is reasonable to deduce that the polymer chain backbone is more planarized when the processing concentrations increase. Combining both observations, we formulate the hypothesis that the more dispersive exciton wave along the polymer chain leads to a greater extent of spatial correlations of energies, which was described in the J-aggregates as shown by Knapp[41] and Knoester.[42] It is worth noting that the spatial correlation function of site energies should be two-dimensional, both along the polymer chain backbone and across the chains in the π-stack, which was demonstrated in P3HT by Spano et al.[10,16,59] Specifically, in the high $M_w$ P3HT, the extent of exciton coherences along (across) the chains are higher (lower) than that of low $M_w$ P3HTs.[16] To accurately account for the exciton coherences of DPP-DTT, rigorous calculations of the two-dimensional correlation functions and a larger range of PL measurements are needed.[16]

# CONCLUSIONS

We demonstrate that the exciton in push-pull polymers is more delocalized in films cast from solutions in which the concentration surpasses the viscosity threshold for gelation. We hypothesize that this phenomenon is attributed to enhanced chain backbone planarization, as indicated by resonance Raman spectroscopy and DFT calculations. Analysis of the absorption and steady-state PL spectra is consistent with a more highly delocalized exciton along the chain backbone, and more chromophores uncoupled to photophysical aggregates are also formed above the gel formation concentration. The contributions to the transient absorption spectra at the timescale of a few picoseconds are likely two-fold: derivative-like spectral line shapes due to accumulating photogenerated charges at the aggregate/non-aggregate interfaces[17] and direct excited-state absorption from the non-aggregate chromophores. We demonstrate the importance of understanding the short-range polymer chain order and excitonic interactions. Manipulating such short-range length scales could further our understanding in preaggregate assembly and favorably accelerate the development of next-generation organic optoelectronics.

# EXPERIMENTAL METHODS

**Sample Preparation.** DPP-DTT ($M_w$ = 290,000 g mol$^{-1}$, dispersity = 2) was purchased from Ossila Limited. For sample preparation, a stock solution of 10 g/L DPP-DTT in chlorobenzene (anhydrous, Sigma-Aldrich) was prepared by heating at 100 °C for around 4 hours, followed by heating at 60 °C overnight. Then solutions with lower concentrations are diluted from the stock solutions. The DPP-DTT thin films are prepared by wire-bar coating the solutions of different concentrations on fused-silica substrates at 56 °C, followed by annealing for 10 min.

**Vis-NIR Absorption Spectroscopy.** The Vis-NIR absorption measurements were performed using the Cary 5000 UV-Vis-NIR spectroscopy.

**Photoluminescence Spectroscopy** The steady-state photoluminescence spectroscopy is performed using the inVia Renishaw Spectrometer in the back-scattering configuration. The samples are illuminated by a 785 nm red laser.

**Ultrafast Transient Absorption Spectroscopy.** The transient absorption measurements are performed using an ultrafast laser system (Pharos Model PH1-20-02-10, Light Conversion). Tunable wavelengths are generated using a laser fundamental of 1030 nm at 100 kHz repetition rate. The integrated transient absorption was measured in a commercial setup (Light Conversion Hera). The wavelengths of the pump can be tuned from 360 to 2600 nm by feeding 10W laser output to a commercial optical parametric amplifier (Orpheus, Light Conversion, Lithuania) while probe beam is generated by sending 2 W to a sapphire crystal to obtain a single-filament white-light continuum in the spectral range of 490-1060 nm. The probe beam was collected by an imaging spectrograph (Shamrock 193i, Andor Technology Ltd., U.K.) coupled with a multichannel detector (256 pixels, 200-1100 nm wavelength range) after transmitting through the sample. All the samples were measured in a homemade vacuum chamber.

**Resonance Raman Spectroscopy.** The resonance Raman spectra are measured using the inVia Renishaw Raman spectrometer, where the samples are excited with 488 nm laser with a back-scattering configuration.

**Quantum Chemistry Calculations.** Density functional theory (DFT) and time-dependent DFT (TDDFT) calculations were performed at the LC-$\omega$HPBE/6-311G(d) level of theory using the Gaussian 16 Rev. A.03 software suite.[60] Empirical gap tuning was performed for the two oligomer geometries following the method of Sun et al.,[61–65] obtaining converged range-separation parameters of $\omega_1 = 0.1295$ (for G1) and $\omega_2 = 0.1216$ (G2). Fol-lowing optimization, vibrational frequency analysis was performed on the oligomer, with vibrational scaling factors of 0.995 (G1) and 0.968 (G2) applied to the calculated Raman frequencies. The Raman activities were converted to intensities consistent with prior literature;[66,67] further details are available in the Supplementary Information. A TDDFT calculation was performed to obtain the excitation wavelength, 439 nm (22780 cm$^{-1}$), corresponding to the wavelength used in the associated experimental spectroscopy.

# Acknowledgement

C.S.A. acknowledges support from the National Science Foundation (Grant DMR-1729737) and from the School of Chemistry and Biochemistry and the College of Science of the Georgia Institute of Technology for startup support. E.R. appreciates support associated with Carl Robert Anderson Chair funds at Lehigh University. The authors also acknowledge the National Science Foundation Grant Nos. 1922111 and 1922174, DMREF: Collaborative Research: Achieving Multicomponent Active Materials through Synergistic Combinatorial, Informatics-enabled Materials Discovery for support. This work was performed in part at the Georgia Tech Institute for Electronics and Nanotechnology, a member of the National Nanotechnology Coordinated Infrastructure (NNCI), which is supported by the National Science Foundation (ECCS-2025462). We further acknowledge the University of Kentucky Center for Computational Sciences and Information Technology Services Research Computing for their fantastic support and collaboration and use of the Lipscomb Compute Cluster and associated research computing resources. The authors thank Hongmo Li, Mark Weber, Marlow Durbin, and Dr. Natalie Stingelin for fruitful discussions on solution-state polymer conformations.

# Supporting Information Available

The Supporting Information is available free of charge at Complete modified FC analysis of the absorption spectra, Gaussian fit of the steady-state PL spectra, DSC thermograms, TA maps measured under the lowest and highest fluences, GSB and ESA signal magnitude dependences upon fluences, double exponential decay fit for the kinetics traces.

# References

(1) Frenkel, J. On the transformation of light into heat in solids. I. Phys. Rev. **1931**, *37*, 17.

(2) Spano, F. C. Excitons in conjugated oligomer aggregates, films, and crystals. Annu. Rev. Phys. Chem. **2006**, *57*, 217–243.

(3) Spano, F. C. Modeling disorder in polymer aggregates: The optical spectroscopy of regioregular poly (3-hexylthiophene) thin films. The Journal of Chemical Physics **2005**, *122*, 234701.

(4) Spano, F. C. Absorption in regio-regular poly (3-hexyl) thiophene thin films: Fermi resonances, interband coupling and disorder. Chemical Physics **2006**, *325*, 22–35.

(5) Clark, J.; Silva, C.; Friend, R. H.; Spano, F. C. Role of Intermolecular Coupling in the Photophysics of Disordered Organic Semiconductors: Aggregate Emission in Regioregular Polythiophene. Phys. Rev. Lett. **2007**, *98*, 206406.

(6) Spano, F. C.; Clark, J.; Silva, C.; Friend, R. H. Determining exciton coherence from the photoluminescence spectral line shape in poly (3-hexylthiophene) thin films. The Journal of Chemical Physics **2009**, *130*, 074904.

(7) Clark, J.; Chang, J.-F.; Spano, F. C.; Friend, R. H.; Silva, C. Determining exciton bandwidth and film microstructure in polythiophene films using linear absorption spectroscopy. Applied Physics Letters **2009**, *94*, 117.

(8) Hestand, N. J.; Spano, F. C. Interference between Coulombic and CT-mediated couplings in molecular aggregates: H-to J-aggregate transformation in perylene-based π-stacks. The Journal of Chemical Physics **2015**, _143_, 244707.

(9) Hestand, N. J.; Spano, F. C. Molecular aggregate photophysics beyond the Kasha model: novel design principles for organic materials. Accounts of chemical research **2017**, _50_, 341–350.

(10) Hestand, N. J.; Spano, F. C. Expanded theory of H-and J-molecular aggregates: the effects of vibronic coupling and intermolecular charge transfer. Chemical Reviews **2018**, _118_, 7069–7163.

(11) Zhong, C.; Bialas, D.; Spano, F. C. Unusual Non-Kasha Photophysical Behavior of Aggregates of Push–Pull Donor–Acceptor Chromophores. The Journal of Physical Chemistry C **2020**, _124_, 2146–2159.

(12) Balooch Qarai, M.; Chang, X.; Spano, F. Vibronic exciton model for low bandgap donor–acceptor polymers. The Journal of Chemical Physics **2020**, _153_, 244901.

(13) Chang, X.; Balooch Qarai, M.; Spano, F. C. HJ-aggregates of donor–acceptor–donor oligomers and polymers. The Journal of Chemical Physics **2021**, _155_, 034905.

(14) Kang, I.; Yun, H.-J.; Chung, D. S.; Kwon, S.-K.; Kim, Y.-H. Record high hole mobility in polymer semiconductors via side-chain engineering. Journal of the American Chemical Society **2013**, _135_, 14896–14899.

(15) Li, J.; Zhao, Y.; Tan, H. S.; Guo, Y.; Di, C.-A.; Yu, G.; Liu, Y.; Lin, M.; Lim, S. H.; Zhou, Y., et al. A stable solution-processed polymer semiconductor with record high-mobility for printed transistors. Scientific Reports **2012**, _2_, 1–9.

(16) Paquin, F.; Yamagata, H.; Hestand, N. J.; Sakowicz, M.; Bérubé, N.; Côté, M.; Reynolds, L. X.; Haque, S. A.; Stingelin, N.; Spano, F. C., et al. Two-dimensional spatial coherence of excitons in semicrystalline polymeric semiconductors: Effect of molecular weight. Physical Review B **2013**, *88*, 155202.

(17) Paquin, F.; Rivnay, J.; Salleo, A.; Stingelin, N.; Silva-Acuña, C. Multi-phase microstructures drive exciton dissociation in neat semicrystalline polymeric semiconductors. Journal of Materials Chemistry C **2015**, *3*, 10715–10722.

(18) Köhler, A.; Hoffmann, S. T.; Bässler, H. An order–disorder transition in the conjugated polymer MEH-PPV. Journal of the American Chemical Society **2012**, *134*, 11594–11601.

(19) Koch, F. P. V.; Rivnay, J.; Foster, S.; Müller, C.; Downing, J. M.; Buchaca-Domingo, E.; Westacott, P.; Yu, L.; Yuan, M.; Baklar, M., et al. The impact of molecular weight on microstructure and charge transport in semicrystalline polymer semiconductors–poly (3-hexylthiophene), a model study. Progress in Polymer Science **2013**, *38*, 1978–1989.

(20) Brinkmann, M.; Rannou, P. Effect of molecular weight on the structure and morphology of oriented thin films of regioregular poly (3-hexylthiophene) grown by directional epitaxial solidification. Advanced Functional Materials **2007**, *17*, 101–108.

(21) Brinkmann, M.; Rannou, P. Molecular weight dependence of chain packing and semicrystalline structure in oriented films of regioregular poly (3-hexylthiophene) revealed by high-resolution transmission electron microscopy. Macromolecules **2009**, *42*, 1125–1130.

(22) Paquin, F.; Latini, G.; Sakowicz, M.; Karsenti, P.-L.; Wang, L.; Beljonne, D.; Stingelin, N.; Silva, C. Charge separation in semicrystalline polymeric semiconductors by photoexcitation: is the mechanism intrinsic or extrinsic? Physical Review Letters **2011**, *106*, 197401.

(23) Reid, O. G.; Malik, J. A. N.; Latini, G.; Dayal, S.; Kopidakis, N.; Silva, C.; Stingelin, N.; Rumbles, G. The influence of solid-state microstructure on the origin and yield of long-lived photogenerated charge in neat semiconducting polymers. Journal of Polymer Science Part B: Polymer Physics **2012**, 50, 27–37.

(24) Noriega, R.; Rivnay, J.; Vandewal, K.; Koch, F. P.; Stingelin, N.; Smith, P.; Toney, M. F.; Salleo, A. A general relationship between disorder, aggregation and charge transport in conjugated polymers. Nature Materials **2013**, 12, 1038–1044.

(25) Sirringhaus, H. 25th anniversary article: organic field-effect transistors: the path beyond amorphous silicon. Advanced Materials **2014**, 26, 1319–1335.

(26) Guo, Z.; Lee, D.; Schaller, R. D.; Zuo, X.; Lee, B.; Luo, T.; Gao, H.; Huang, L. Relationship between interchain interaction, exciton delocalization, and charge separation in low-bandgap copolymer blends. Journal of the American Chemical Society **2014**, 136, 10024–10032.

(27) Steyrleuthner, R.; Schubert, M.; Howard, I.; Klaumuünzer, B.; Schilling, K.; Chen, Z.; Saalfrank, P.; Laquai, F.; Facchetti, A.; Neher, D. Aggregation in a high-mobility n-type low-bandgap copolymer with implications on semicrystalline morphology. Journal of the American Chemical Society **2012**, 134, 18303–18317.

(28) Gélinas, S.; Kirkpatrick, J.; Howard, I. A.; Johnson, K.; Wilson, M. W.; Pace, G.; Friend, R. H.; Silva, C. Recombination dynamics of charge pairs in a push–pull polyfluorene-derivative. The Journal of Physical Chemistry B **2013**, 117, 4649–4653.

(29) Pop, F.; Lewis, W.; Amabilino, D. B. Solid state supramolecular structure of diketopyrrolopyrrole chromophores: correlating stacking geometry with visible light absorption. CrystEngComm **2016**, 18, 8933–8943.

(30) Chaudhari, S. R.; Griffin, J. M.; Broch, K.; Lesage, A.; Lemaur, V.; Dudenko, D.; Olivier, Y.; Sirringhaus, H.; Emsley, L.; Grey, C. P. Donor–acceptor stacking arrangements in bulk and thin-film high-mobility conjugated polymers characterized using molecular modelling and MAS and surface-enhanced solid-state NMR spectroscopy. Chemical Science **2017**, *8*, 3126–3136.

(31) McRae, E. G.; Kasha, M. Enhancement of phosphorescence ability upon aggregation of dye molecules. The Journal of Chemical Physics **1958**, *28*, 721–722.

(32) Yan, M.; Rothberg, L.; Papadimitrakopoulos, F.; Galvin, M.; Miller, T. Spatially indirect excitons as primary photoexcitations in conjugated polymers. Physical Review Letters **1994**, *72*, 1104.

(33) Cacialli, F.; Wilson, J. S.; Michels, J. J.; Daniel, C.; Silva, C.; Friend, R. H.; Severin, N.; Samorì, P.; Rabe, J. P.; O'Connell, M. J., et al. Cyclodextrin-threaded conjugated polyrotaxanes as insulated molecular wires with reduced interstrand interactions. Nature Materials **2002**, *1*, 160–164.

(34) Petrozza, A.; Brovelli, S.; Michels, J. J.; Anderson, H. L.; Friend, R. H.; Silva, C.; Cacialli, F. Control of Rapid Formation of Interchain Excited States in Sugar-Threaded Supramolecular Wires. Advanced Materials **2008**, *20*, 3218–3223.

(35) Mróz, M. M.; Perissinotto, S.; Virgili, T.; Gigli, G.; Salerno, M.; Frampton, M. J.; Sforazzini, G.; Anderson, H. L.; Lanzani, G. Laser action from a sugar-threaded polyrotaxane. Applied Physics Letters **2009**, *95*, 031108.

(36) Cabanillas-Gonzalez, J.; Grancini, G.; Lanzani, G. Pump-probe spectroscopy in organic semiconductors: monitoring fundamental processes of relevance in optoelectronics. Advanced Materials **2011**, *23*, 5468–5485.

(37) Mróz, M. M.; Sforazzini, G.; Zhong, Y.; Wong, K. S.; Anderson, H. L.; Lanzani, G.; Cabanillas-Gonzalez, J. Amplified spontaneous emission in conjugated polyrotaxanes under quasi-cw pumping. Advanced Materials **2013**, *25*, 4347–4351.

(38) Wang, Z.; Gao, M.; He, C.; Shi, W.; Deng, Y.; Han, Y.; Ye, L.; Geng, Y. Unraveling the Molar Mass Dependence of Shearing-Induced Aggregation Structure of a High-Mobility Polymer Semiconductor. Advanced Materials **2022**, 2108255.

(39) Venkatesh, R.; Zheng, Y.; Liu, A. L.; Zhao, H.; Silva, C.; Takacs, C. J.; Grover, M.; Meredith, C.; Reichmanis, E. Overlap concentration generates optimum device performance for DPP-based conjugated polymers. Organic Electronics **2023**, 106779.

(40) Venkatesh, R.; Zheng, Y.; Viersen, C.; Liu, A.; Silva, C.; Grover, M.; Reichmanis, E. Data Science Guided Experiments Identify Conjugated Polymer Solution Concentration as a Key Parameter in Device Performance. ACS Materials Letters **2021**, *3*, 1321–1327.

(41) Knapp, E. Lineshapes of molecular aggregates, exchange narrowing and intersite correlation. Chemical Physics **1984**, *85*, 73–82.

(42) Knoester, J. Nonlinear optical line shapes of disordered molecular aggregates: motional narrowing and the effect of intersite correlations. The Journal of Chemical Physics **1993**, *99*, 8466–8479.

(43) Li, M.; Balawi, A. H.; Leenaers, P. J.; Ning, L.; Heintges, G. H.; Marszalek, T.; Pisula, W.; Wienk, M. M.; Meskers, S. C.; Yi, Y., et al. Impact of polymorphism on the optoelectronic properties of a low-bandgap semiconducting polymer. Nature Communications **2019**, *10*, 1–11.

(44) Peng, Z.; Ye, L.; Ade, H. Understanding, quantifying, and controlling the molecular ordering of semiconducting polymers: from novices to experts and amorphous to perfect crystals. Materials Horizons **2022**, *9*, 577–606.

(45) Yamagata, H.; Spano, F. C. Interplay between intrachain and interchain interactions in semiconducting polymer assemblies: The HJ-aggregate model. The Journal of Chemical Physics **2012**, *136*, 184901.

(46) Vezie, M. S.; Few, S.; Meager, I.; Pieridou, G.; Dörling, B.; Ashraf, R. S.; Goñi, A. R.; Bronstein, H.; McCulloch, I.; Hayes, S. C., et al. Exploring the origin of high optical absorption in conjugated polymers. Nature Materials **2016**, *15*, 746–753.

(47) Weiser, G. Stark effect of one-dimensional Wannier excitons in polydiacetylene single crystals. Physical Review B **1992**, *45*, 14076.

(48) Liess, M.; Jeglinski, S.; Vardeny, Z.; Ozaki, M.; Yoshino, K.; Ding, Y.; Barton, T. Electroabsorption spectroscopy of luminescent and nonluminescent $\pi$-conjugated polymers. Physical Review B **1997**, *56*, 15712.

(49) Wood, S.; Wade, J.; Shahid, M.; Collado-Fregoso, E.; Bradley, D. D.; Durrant, J. R.; Heeney, M.; Kim, J.-S. Natures of optical absorption transitions and excitation energy dependent photostability of diketopyrrolopyrrole (DPP)-based photovoltaic copolymers. Energy & Environmental Science **2015**, *8*, 3222–3232.

(50) Banerji, N.; Gagnon, E.; Morgantini, P.-Y.; Valouch, S.; Mohebbi, A. R.; Seo, J.-H.; Leclerc, M.; Heeger, A. J. Breaking down the problem: optical transitions, electronic structure, and photoconductivity in conjugated polymer PCDTBT and in its separate building blocks. The Journal of Physical Chemistry C **2012**, *116*, 11456–11469.

(51) Li, Y.; Tatum, W. K.; Onorato, J. W.; Zhang, Y.; Luscombe, C. K. Low elastic modulus and high charge mobility of low-crystallinity indacenodithiophene-based semiconducting polymers for potential applications in stretchable electronics. Macromolecules **2018**, *51*, 6352–6358.

(52) Rolczynski, B. S.; Szarko, J. M.; Son, H. J.; Liang, Y.; Yu, L.; Chen, L. X. Ultrafast intramolecular exciton splitting dynamics in isolated low-band-gap polymers and their implications in photovoltaic materials design. Journal of the American Chemical Society **2012**, *134*, 4142–4152.

(53) Tautz, R.; Da Como, E.; Limmer, T.; Feldmann, J.; Egelhaaf, H.-J.; Von Hauff, E.; Lemaur, V.; Beljonne, D.; Yilmaz, S.; Dumsch, I., et al. Structural correlations in the generation of polaron pairs in low-bandgap polymers for photovoltaics. Nature Communications **2012**, *3*, 1–8.

(54) Yamagata, H.; Pochas, C. M.; Spano, F. C. Designing J-and H-aggregates through wave function overlap engineering: applications to poly (3-hexylthiophene). The Journal of Physical Chemistry B **2012**, *116*, 14494–14503.

(55) Lo, C. K.; Gautam, B. R.; Selter, P.; Zheng, Z.; Oosterhout, S. D.; Constantinou, I.; Knitsch, R.; Wolfe, R. M.; Yi, X.; Brédas, J.-L., et al. Every atom counts: Elucidating the fundamental impact of structural change in conjugated polymers for organic photovoltaics. Chemistry of Materials **2018**, *30*, 2995–3009.

(56) Seifrid, M.; Reddy, G.; Chmelka, B. F.; Bazan, G. C. Insight into the structures and dynamics of organic semiconductors through solid-state NMR spectroscopy. Nature Reviews Materials **2020**, *5*, 910–930.

(57) Melnyk, A.; Junk, M. J.; McGehee, M. D.; Chmelka, B. F.; Hansen, M. R.; Andrienko, D. Macroscopic structural compositions of $\pi$-conjugated polymers: combined insights from solid-state NMR and molecular dynamics simulations. The Journal of Physical Chemistry Letters **2017**, *8*, 4155–4160.

(58) McCulloch, I.; Heeney, M.; Bailey, C.; Genevicius, K.; MacDonald, I.; Shkunov, M.; Sparrowe, D.; Tierney, S.; Wagner, R.; Zhang, W., et al. Liquid-crystalline semiconducting polymers with high charge-carrier mobility. Nature Materials **2006**, *5*, 328–333.

(59) Spano, F. C.; Silva, C. H-and J-aggregate behavior in polymeric semiconductors. Annual Review of Physical Chemistry **2014**, *65*, 477–500.

(60) Frisch, M. J. et al. Gaussian~16 Revision A.03. 2016; Gaussian Inc. Wallingford CT.

(61) Sun, H.; Ryno, S.; Zhong, C.; Ravva, M. K.; Sun, Z.; Körzdörfer, T.; Brédas, J.-L. Ionization Energies, Electron Affinities, and Polarization Energies of Organic Molecular Crystals: Quantitative Estimations from a Polarizable Continuum Model (PCM)-Tuned Range-Separated Density Functional Approach. Journal of Chemical Theory and Computation **2016**, *12*, 2906–2916.

(62) Henderson, T. M.; Izmaylov, A. F.; Scalmani, G.; Scuseria, G. E. Can short-range hybrids describe long-range-dependent properties? The Journal of Chemical Physics **2009**, *131*, 044108.

(63) Vreven, T.; Frisch, M. J.; Kudin, K. N.; Schlegel, H. B.; Morokuma, K. Geometry optimization with QM/MM methods II: Explicit quadratic coupling. Molecular Physics **2006**, *104*, 701–714.

(64) Vydrov, O. A.; Scuseria, G. E. Assessment of a long-range corrected hybrid functional. The Journal of Chemical Physics **2006**, *125*, 234109.

(65) Vydrov, O. A.; Scuseria, G. E.; Perdew, J. P. Tests of functionals for systems with fractional electron number. The Journal of Chemical Physics **2007**, *126*, 154109.

(66) Polavarapu, P. L. Ab initio vibrational Raman and Raman optical activity spectra. The Journal of Physical Chemistry **1990**, *94*, 8106–8112.

(67) Keresztury, G.; Holly, S.; Besenyei, G.; Varga, J.; Wang, A.; Durig, J. Vibrational spectra of monothiocarbamates-II. IR and Raman spectra, vibrational assignment, conformational analysis and ab initio calculations of S-methyl-N,N-dimethylthiocarbamate. Spectrochimica Acta Part A: Molecular Spectroscopy **1993**, *49*, 2007–2026.

# Supporting Information: Chain Conformation and Exciton Delocalization in a Push-Pull Conjugated Polymer

Yulong Zheng,[†] Rahul Venkatesh,[‡] Connor P. Callaway,[¶] Campbell Viersen,[†] Kehinde H. Fagbohungbe,[¶] Aaron L. Liu,[‡] Chad Risko,[¶] Elsa Reichmanis,[§] and Carlos Silva-Acuña[*,†,‖,⊥]

[†]School of Chemistry and Biochemistry, Georgia Institute of Technology, 901 Atlantic Drive, Atlanta GA 30332, United States

[‡]School of Chemical and Biomolecular Engineering, Georgia Institute of Technology, 311 Ferst Drive NW, Atlanta GA 30332, United States

[¶]Department of Chemistry and Center for Applied Energy Research, University of Kentucky, Lexington, Kentucky 40506, United-States

[§]Department of Chemical & Biomolecular Engineering, Lehigh University, 111 Research Drive, Bethlehem PA 18015, United States

[‖]School of Physics, Georgia Institute of Technology, 837 State Street, Atlanta GA 30332, United States

[⊥]School of Materials Science and Engineering, Georgia Institute of Technology, 771 Ferst Drive NW, Atlanta GA 30332, United States

E-mail: carlos.silva@gatech.edu

# Absorption and Steady-state PL

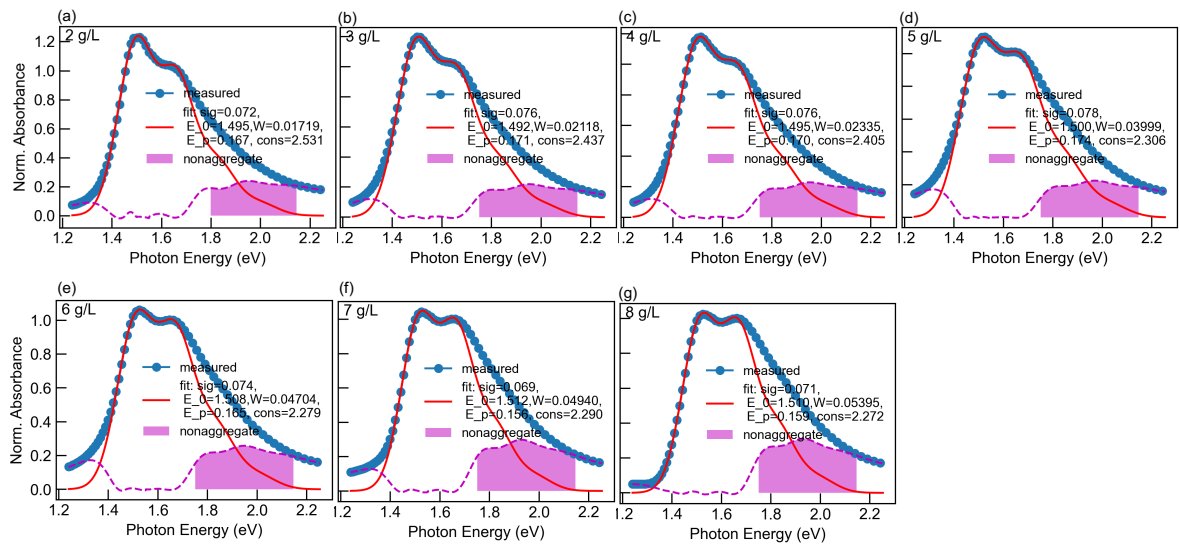

**Figure S1:** (a)-(g) Modified FC fit for all absorption spectra of the thin films prepared from different concentrations. The parameters are shown in the main article Table 1. Except for the constant, all other parameters are displayed in units of eV.

The modified Franck-Condon analysis of the samples with different concentrations is shown for each concentration in Figure S1. The complete set of parameters acquired is shown in TableS1. The *effective* non-aggregate spectra are acquired by subtracting the simulated aggregate spectra from the measured one. It is worth mentioning that the absorption spectra of push-pull polymers have a camel-back feature, where the absorbance between the high and low energy band is not completely zero due to the strong FE/CT mixing.[1] The oscillator strength ratios of the non-aggregate and aggregates absorption spectra are plotted in Figure S1(f).

The PL spectra are fitted with a single Gaussian distribution as shown in Figure S2. By subtracting the measurements with the Gaussian fit, a high-energy peak is observed when the samples are prepared from concentrations surpassing the critical chain overlap point.

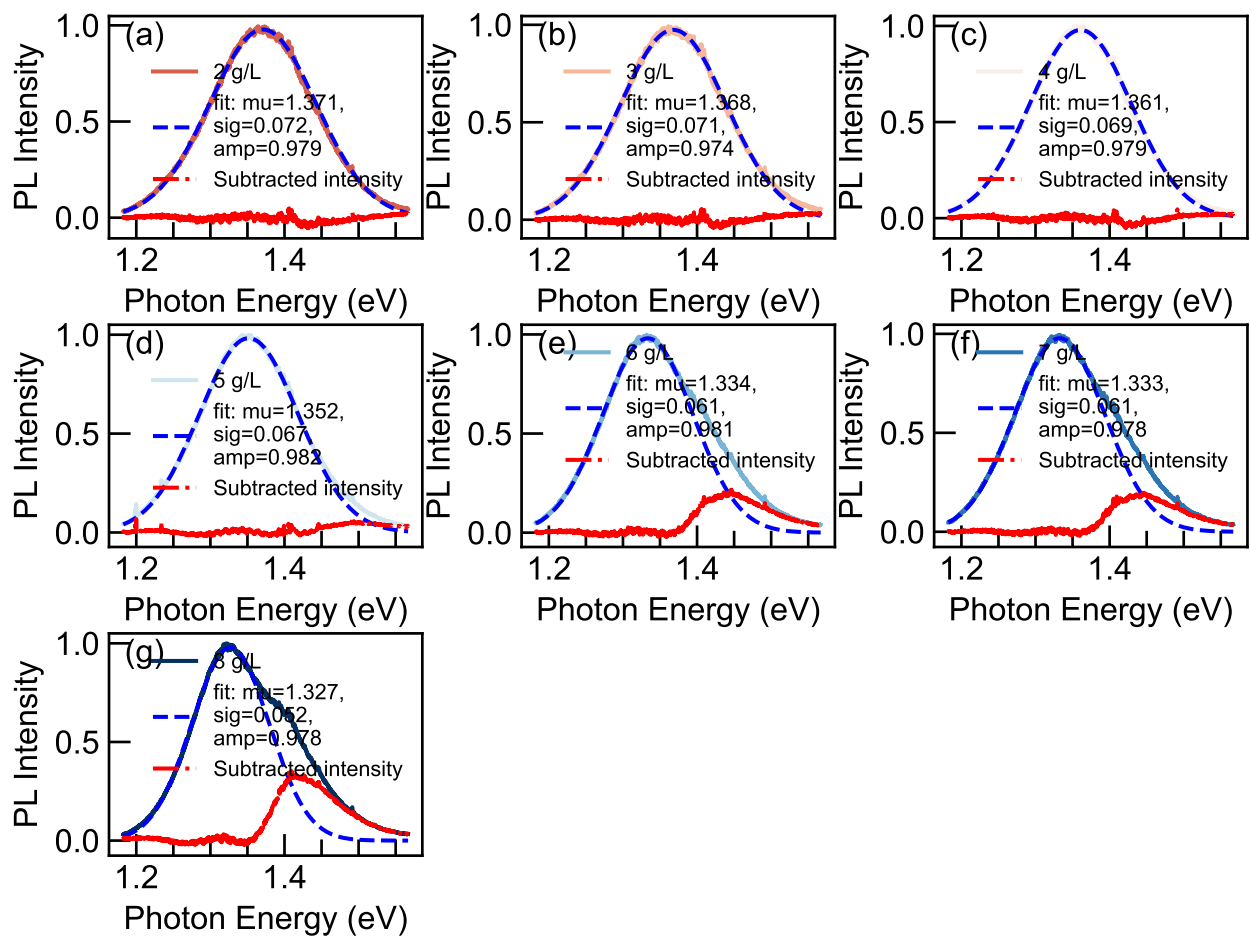

**Figure S2:** The Gaussian fit (blue dashed line) for PL spectra of thin film samples prepared from each concentration. The high energy shoulder is indicated by the red line.

Table S1: Parameters Acquired from Modified FC Fit

| Conc. (g/L) | $\sigma$ (meV) | $E_0$ (meV) | W (meV) | $E_p$ (meV) | Cons. |
|---|---|---|---|---|---|
| 2 | 71.9 ± 0.3 | 1495 ± 0.3 | 17.2 ± 0.6 | 167.0 ± 0.6 | 2.531 ± 0.006 |
| 3 | 75.7 ± 0.2 | 1492 ± 0.1 | 21.2 ± 0.3 | 171.0 ± 0.5 | 2.437 ± 0.002 |
| 4 | 75.9 ± 0.2 | 1495 ± 0.1 | 23.3 ± 0.3 | 170.1 ± 0.4 | 2.405 ± 0.002 |
| 5 | 78.5 ± 0.1 | 1500 ± 0.1 | 40.0 ± 0.2 | 173.9 ± 0.3 | 2.306 ± 0.001 |
| 6 | 73.9 ± 0.2 | 1508 ± 0.2 | 47.0 ± 0.3 | 164.7 ± 0.4 | 2.279 ± 0.002 |
| 7 | 69.4 ± 0.2 | 1512 ± 0.2 | 49.4 ± 0.3 | 156.5 ± 0.4 | 2.290 ± 0.003 |
| 8 | 70.7 ± 0.2 | 1511 ± 0.2 | 53.9 ± 0.4 | 159.4 ± 0.5 | 2.272 ± 0.003 |

From left to right are the widths of Gaussian inhomogeneous distribution used to describe the energetic disorder, $\sigma$; the energy of the 0-0 vibronic transition, $E_0$; exciton bandwidth, W; the energy of the vibrational mode coupled to the electronic transition, $E_p$ and the fit constant.

## Differential Scanning Calorimetry

The differential scanning calorimetry is conducted with TA Instruments DSC250. The thermograph is shown in Figure S3. With a cooling and heating rate of 20 °C/min, no meaningful differences are observed with samples prepared from 2 to 8 g/L. All heating curves display melting peaks at 375±2 °C with endothermic enthalpy of 32±4 °C.

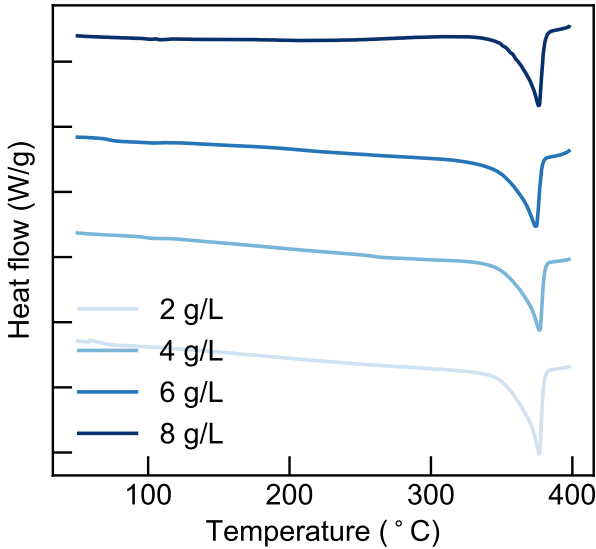

**Figure S3:** First heating curves of drop-cast films prepared from concentrations of 2, 4, 6 and 8 g/L with endo-down heating flow.

Table S2: DPP-DTT Thermal Properties

| Conc. (g/L) | enthalpy (J/g) | onset temp. (°C) | peak temp. (°C) |
|---|---|---|---|
| 2 | 32 | 364 | 377 |
| 4 | 29 | 362 | 377 |
| 6 | 36 | 356 | 374 |
| 8 | 29 | 362 | 376 |

The first heating DSC thermogram for DPP-DTT thin films precipitated from solution concentrations of 2, 4, 6 and 8 g/L. The heating and cooling rate is 20 °C/min.

# Transient absorption

The transient absorption spectra measured under the lowest (Fig. S4) and highest pump fluences (Figure S6) that are feasible for clear signals, respectively. The fluences are displayed in each spectrum. We also took the temporal cuts at 1, 10, 100 and 800 time delays pumped at low fluences as shown in Figure S5. Due to the relative low fluences, signal-to-noise ratio at time delays beyond 100 ps are relatively low. It might be one of the reasons for the unusual feature in Figure 3 in the main article.

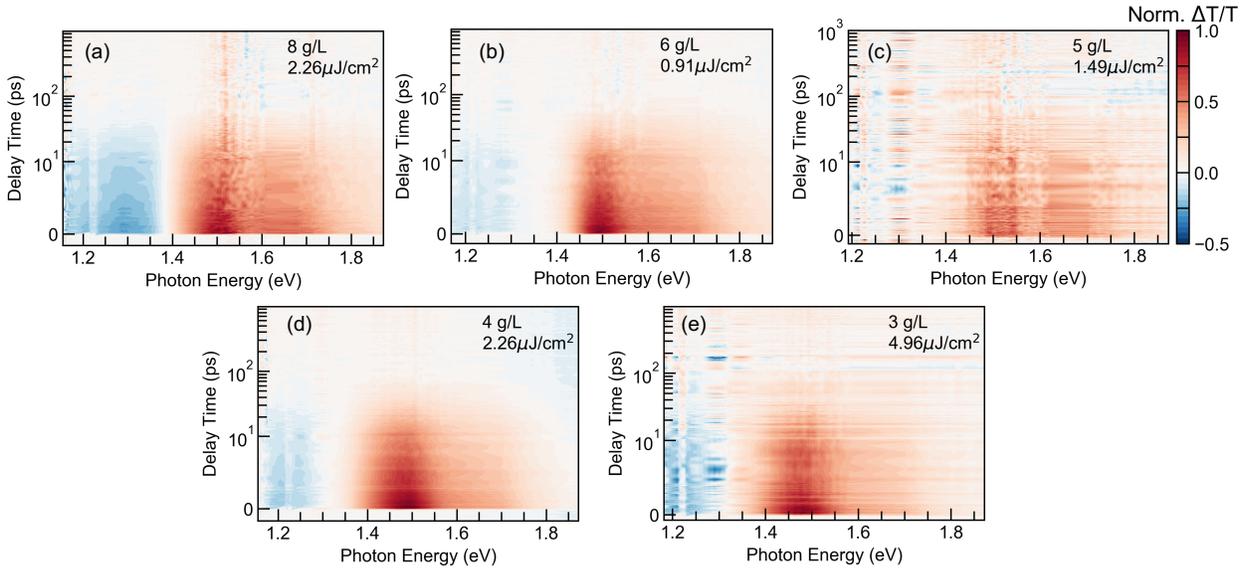

**Figure S4:** Normalized transient absorption time-energy map collected at the lowest pump fluences for 3, 4, 5, 6 and 8 g/L. The fluences are indicated in the annotation.

For the purpose of demonstration here, the transient absorption decays at 750 and 950

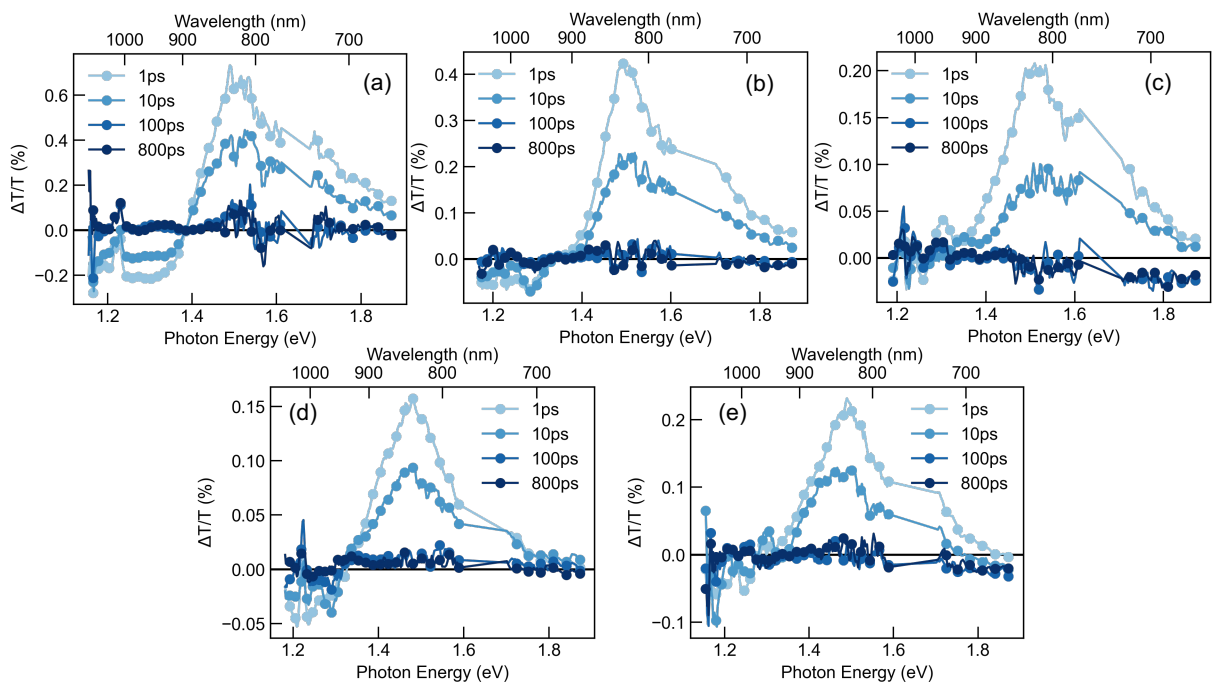

**Figure S5:** The spectra cuts taken at varying time delays. (a), (b), (c), (d), and (e) are the spectra of 8, 6, 5, 4, and 3 g/L, respectively. The corresponding fluences are shown in Figure S4

nm for 8 g/L are displayed in Figure S7. Although not directly shown here, the samples of other concentrations show similar dynamics. At the fluence of 2.2 $\mu$J/cm$^2$, the apparent monoexponential lifetime is around 18 ps, where we assume the exciton-exciton annihilation is not significant at such low fluence.

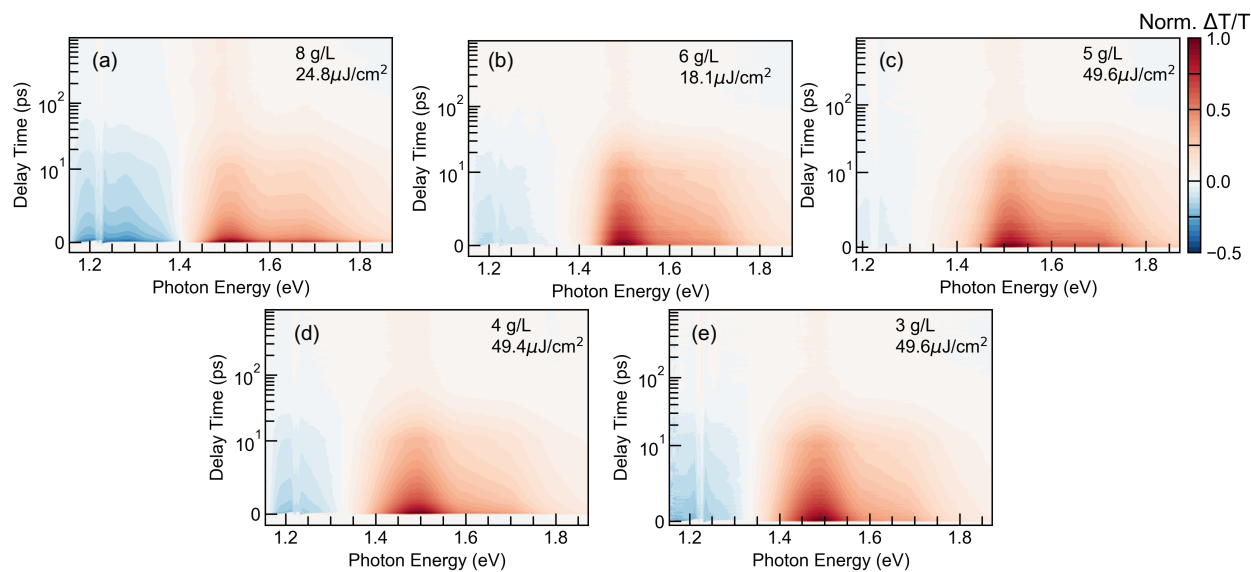

**Figure S6:** Normalized transient absorption time-energy map collected at the highest pump fluences for 3, 4, 5, 6 and 8 g/L. The fluences are indicated in the annotation.

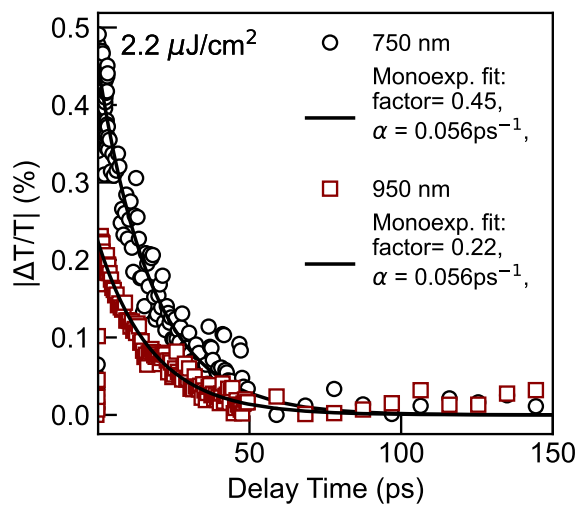

**Figure S7:** The TA decays at 750 nm (black circle) and 950 nm (brown square), respectively. in the 8g/L film. The solid lines display the monoexponential fit.

# Quantum Chemistry Calculations

Density functional theory (DFT) and time-dependent DFT (TDDFT) calculations were performed using the Gaussian 16 Rev. A.03 software suite.[2] A DPP-DTT trimer, terminated at both ends with thiophene-flanked DPP units for symmetry, was optimized at the LC-$\omega$HPBE/6-311G(d) level of theory. Empirical gap tuning was performed for the two oligomer geometries shown in Figure S8 following the method of Sun et al.,[3–7] obtaining converged range-separation parameters of $\omega_1 = 0.1295$ for geometry 1 (G1) and $\omega_2 = 0.1216$ for geometry 2 (G2). Alkyl side chains on DPP units were truncated to methyl units to reduce computational cost. Following optimization, vibrational frequency analysis was performed on the oligomer to verify that the obtained geometry was an energy minimum, as well as to obtain IR and Raman spectra and vibrational modes. Vibrational scaling factors of 0.995 (G1) and 0.968 (G2) were applied to the calculated Raman frequencies. The Raman activities were converted to intensities $I_i$ using the scattering equation

$$I_i = \frac{(\nu_0 - \nu_i)^4 S_i}{\nu_i [1 - \exp(-\frac{hc\nu_i}{kT})]}$$

where $\nu_0$ is the excitation wavelength; $\nu_i$ and $S_i$ are the spatial frequency and Raman activity of vibrational mode $i$; and $h$, $c$, and $k$ are Planck's constant, the speed of light, and Boltzmann's constant, respectively.[8,9] The temperature was taken to be 298.15 K. A TDDFT calculation was performed to obtain the excitation wavelength, 439 nm (22780 cm$^{-1}$), corresponding to the wavelength used in the associated experimental spectroscopy.

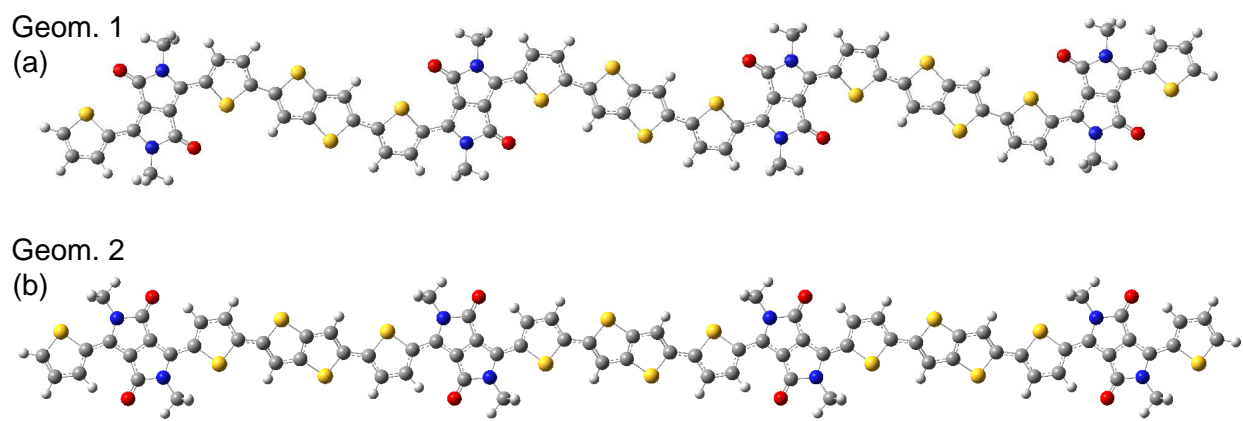

**Figure S8:** DFT-optimized molecular structures of (a) geometry 1 (G1) and (b) geometry 2 (G2). Notice that in G1, the nitrogen atom in DPP unit is close to the sulfur atom in the neighboring thiophene unit, while in G2, the nitrogen atom is in the vicinity of the hydrogen atom in the neighboring thiophene unit.

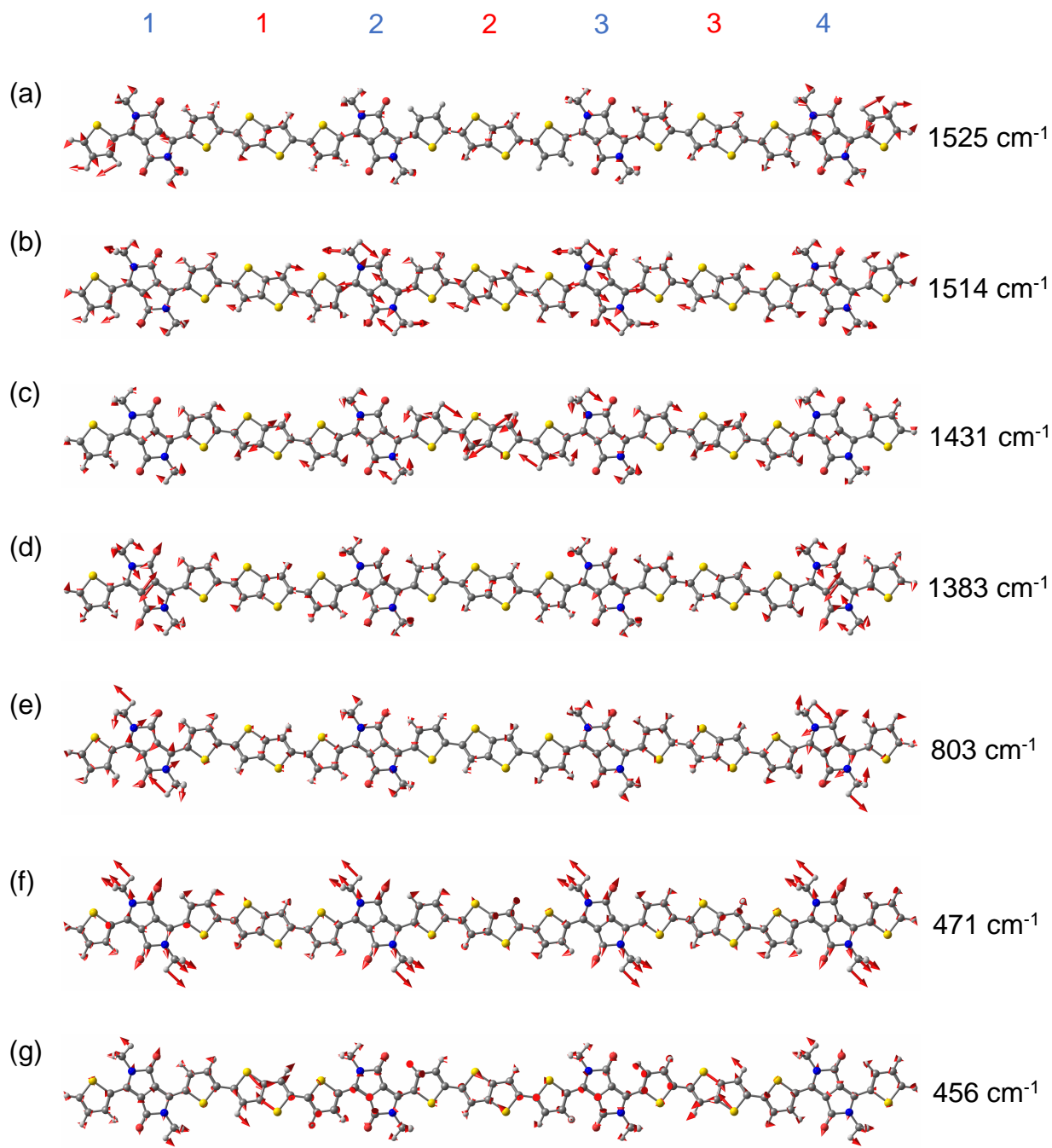

**Figure S9:** The vector diagrams for the modes of interest in geometry 2, corresponding to the geometry 1 modes shown in Figure 6 in the main article.

# References

(1) Banerji, N.; Gagnon, E.; Morgantini, P.-Y.; Valouch, S.; Mohebbi, A. R.; Seo, J.-H.; Leclerc, M.; Heeger, A. J. Breaking down the problem: optical transitions, electronic structure, and photoconductivity in conjugated polymer PCDTBT and in its separate building blocks. *The Journal of Physical Chemistry C* **2012**, *116*, 11456–11469.

(2) Frisch, M. J.; Trucks, G. W.; Schlegel, H. B.; Scuseria, G. E.; Robb, M. A.; Cheeseman, J. R.; Scalmani, G.; Barone, V.; Petersson, G. A.; Nakatsuji, H. et al. Gaussian~16 Revision A.03. 2016; Gaussian Inc. Wallingford CT.

(3) Sun, H.; Ryno, S.; Zhong, C.; Ravva, M. K.; Sun, Z.; Körzdörfer, T.; Brédas, J.-L. Ionization Energies, Electron Affinities, and Polarization Energies of Organic Molecular Crystals: Quantitative Estimations from a Polarizable Continuum Model (PCM)-Tuned Range-Separated Density Functional Approach. *Journal of Chemical Theory and Computation* **2016**, *12*, 2906–2916.

(4) Henderson, T. M.; Izmaylov, A. F.; Scalmani, G.; Scuseria, G. E. Can short-range hybrids describe long-range-dependent properties? *The Journal of Chemical Physics* **2009**, *131*, 044108.

(5) Vreven, T.; Frisch, M. J.; Kudin, K. N.; Schlegel, H. B.; Morokuma, K. Geometry optimization with QM/MM methods II: Explicit quadratic coupling. *Molecular Physics* **2006**, *104*, 701–714.

(6) Vydrov, O. A.; Scuseria, G. E. Assessment of a long-range corrected hybrid functional. *The Journal of Chemical Physics* **2006**, *125*, 234109.

(7) Vydrov, O. A.; Scuseria, G. E.; Perdew, J. P. Tests of functionals for systems with fractional electron number. *The Journal of Chemical Physics* **2007**, *126*, 154109.

# TOC Graphic

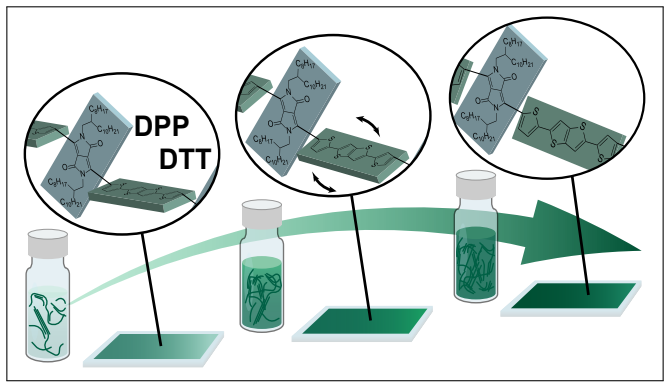